\documentclass[aps,pre,reprint,showpacs,showkeys,noeprint]{revtex4-1} 
\usepackage[utf8]{inputenc}
\usepackage[T1]{fontenc}
\usepackage{amsmath,amssymb,mathtools}
\usepackage{txfonts}

\usepackage{mathrsfs}

\usepackage{graphicx,grffile}
\graphicspath{{/}}

\usepackage{textcomp}
\usepackage{booktabs,tabularx,dcolumn}
\newcolumntype{d}[1]{D{.}{.}{#1}}

\usepackage{url,hyperref}
\usepackage[usenames,dvipsnames]{xcolor}
\hypersetup{colorlinks=true, linkcolor=BrickRed, urlcolor=blue!50!black, citecolor=blue!50!black}

\usepackage[capitalize]{cleveref}
\crefrangelabelformat{equation}{(#3#1#4)$-$(#5#2#6)}

\renewcommand\geq{\geqslant}
\renewcommand\rho\varrho
\renewcommand\vec[1]{\textrm{\bfseries #1}}

\usepackage[normalem]{ulem}

\begin{document}
\title{Dynamic Scaling and Aging in Hydrodynamic Coarsening of Vapor Bubbles in a Lennard-Jones Fluid}

\author{Sutapa Roy}
\email{sutapa.roy@hyderabad.bits-pilani.ac.in}
\affiliation{Department of Physics, Birla Institute of Technology and Science, Pilani, Hyderabad Campus, Jawahar Nagar, Kapra Mandal, Medchal District, Telangana 500078, India}

\date{\today}

\begin{abstract}
We investigate the coarsening and aging dynamics of vapor bubbles during liquid-vapor phase separation in a three-dimensional single-component Lennard-Jones fluid using large-scale molecular dynamics simulations. 
Following a temperature quench from a homogeneous high-temperature state into the two-phase coexistence region, vapor bubbles nucleate and grow in the liquid background. 
The characteristic bubble size follows a power-law growth, $\ell(t)\sim t^\alpha$, with a growth exponent $\alpha \simeq 2/3$, characteristic of inertial hydrodynamic coarsening. 
We also investigate the two-time autocorrelation function $C_{ag}(t,t_w)$ which follows waiting-time dependent relaxation, revealing aging. 
Furthermore, data at different waiting times collapse onto a scaling function when plotted as a function of the length ratio $\ell(t)/\ell(t_w)$, demonstrating dynamical scaling during aging. 
The corresponding scaling function exhibits an algebraic decay with an exponent $\lambda \simeq 3$. 
Interestingly, bubble and droplet growth exhibit distinctly different domain-growth and aging exponents despite both being disconnected morphologies. 
\end{abstract}

\maketitle

\section{Introduction}
Phase separation is a fundamental non-equilibrium process during which an initially homogeneous system evolves toward a state containing distinct phases \cite{puri2024, puri2009kinetics, bray2002theory, onuki2002phase}. 
It is observed in a wide variety of systems including liquid crystals \cite{puri-nigel}, polymers \cite{jiang2013, voit2005}, chemically active matter \cite{nalina2024}, biological systems \cite{condensate2025, su2024kinetic, wangcancer2025, tong2022,  zhang2020liquid}, and holds technological applications, e.g., inkjets \cite{inkjet2022}. 
Following a sudden temperature quench below the critical point of a second-order phase transition \cite{demixing}, a system falls out of equilibrium and moves towards the new equilibrium. 
The subsequent evolution is characterized by the formation and growth of domains of like particles  \cite{puri2009kinetics} as the system progressively reduces its interfacial free energy. 
This non-equilibrium process is known as coarsening or phase separation dynamics. 

Apart from isothermal, bulk systems, phase separation has also been studied in the presence of confinement \cite{Wu2026Surface, Puri2005}. 
External shear can alter the phase separation behavior \cite{zhang2008, onuki1986}, elongate the domains \cite{Gonnella_2020}, or can even arrest the domain growth \cite{gonella2025}. 
Shear is also known to accelerate aggregation of colloidal particles \cite{zaccone2009}.
Domain growth near a surface depends on the interplay between bulk demixing, surface preference for a component, and the finite film thickness \cite{binder2010}. 
Stirring can suppress phase separation with the composition converging exponentially towards a spatially homogeneous mixed state \cite{feng2020}.
Presence of inhomogeneous temperature fields makes the coarsening mechanism particularly interesting. \cite{lamorgese2011, araki2004}. 
The directions of the temperature gradient and the initial concentration gradient have been shown \cite{Hong_2010} to strongly influence the resulting anisotropic morphology.
Directional quenches \cite{kurita2022, kurita2017, furukawa-physica1992} give rise to patterns which are different from bulk and can be used to create controlled domain morphology. 
A moving control-parameter front can select the size of alternating domains left behind it \cite{foard-pre2009} and steer the growing stripes \cite{prl2006biro}. 
By tuning the velocities of the moving cooling source and thermal front propagation, one can engineer the desired domain morphology \cite{klp2026}. 
Lattice-Boltzmann studies showed that the speed of a moving front and mixture composition can select domain shape, orientation, and size \cite{LBpre2012}. 
Traveling periodic forcing can organize phase-separated domains and drag them along \cite{wave2009}. 
Shear flow and nonuniform cooling interact to produce patterns more complex than either mechanism produces separately \cite{bertei2021}.
Beyond conventional phase separation, interaction-driven self-organization can also produce complex spatial structures in a broad spectrum of systems \cite{laradji2025, alois2019}.
In addition to experimental methods and analytical theory, a number of simulation methods are applied to study this phenomena, viz., Lattice-Boltzmann simulation \cite{furtado2006}, Monte-Carlo \cite{das2025, Zausch_2010, landau2009guide}, phase-field modeling \cite{mino2015}, molecular dynamics simulations \cite{roynalina, royepl}, etc. 

Domain growth is known to strongly depend on the density/concentration of the fluid, order parameter symmetry, order parameter conservation, system dimension, and the role of transport processes \cite{puri2009kinetics, binder1991spinodal}. 
In general, the average domain size $\ell(t)$ grows with time in a power-law manner as \cite{puri2009kinetics, bray2002theory}
\begin{equation} \label{growth}
\ell(t) \sim t^{\alpha},
\end{equation}
where, $\alpha$ is the growth exponent. 
This reflects self-similar nature of domain growth. 
For solids, mass transport is dominated by diffusion and the  Lifshitz-Slyozov-Wagner (LSW) theory predicts a characteristic length scale that grows with $\alpha=1/3$ \cite{binder1991spinodal, lifshitz1961kinetics}.
In fluids, however, the order parameter gets coupled to momentum and hydrodynamic flow and substantially modifies the coarsening kinetics \cite{laradji1996molecular, roy2013effects}. 
As suggested via Siggia, tube-like percolating domain structures create a pressure gradient due to interfacial tension and leads to a much faster fluid flow through them. 
Balancing of interfacial and viscous stresses leads to a linear growth exponent, $\ell(t)\sim t$ \cite{edsiggia1979}. 
This is known as the viscous hydrodynamic growth. 
On the other hand, at very late times when domains get large enough inertia becomes important. 
Inertial hydrodynamic effects produce a very different behavior $\ell(t) \sim t^{2/3}$ \cite{furukawa1987}. 

Domain morphology depends on the quench concentration as well. 
For a quench at the critical concentration, percolating structures are formed. 
On the other hand, for an off-critical quench droplets nucleate and grow \cite{roy2013dynamics, roy2013effects, binder2016} with an exponent $\alpha=1/3$ following a hydrodynamic mechanism \cite{binder1977theory, roycoalescence} which is different from LSW. 
Here, diffusion-controlled binary collisions occur between droplets. 
The collision/coalescence rate can be expressed as
\begin{equation} \label{BSeq}
    \frac{dn}{dt}=-C D\ell\,n^2,
\end{equation}
where $D$ is the droplet diffusion coefficient and $C$ is a constant.
For a fixed minority-phase volume fraction $\phi$,
$n\ell^3\sim \phi$, so that $n\sim \ell^{-3}$. 
Using the Stokes–Einstein-Sutherland relation for a spherical droplet
\begin{equation}
    D_\ell=\frac{k_BT}{6\pi\eta\ell},
\end{equation}
$k_B$, $T$, $\eta$ being the Boltzmann constant, fluid temperature and viscosity, respectively, one obtains
\begin{equation}
    D_\ell\ell=\text{constant}.
\end{equation}
Consequently, the rate equation becomes
\begin{equation}
    \frac{dn}{dt}\sim-n^2.
\end{equation}
Integration gives $n(t)\sim t^{-1}$. 
But because $n\sim\ell^{-3}$, 
\begin{equation}
    \ell(t)\sim t^{1/3}.
\end{equation}
This growth mechanism is also known as the Binder–Stauffer (BS) mechanism of droplet coarsening. 
Even though the exponent is $1/3$, the mechanism is not same as the LSW evaporation–condensation mechanism. 
LSW involves molecular evaporation/condensation. 
On the other hand, in the BS mechanism, droplets undergo Brownian translational motion.
When two droplets encounter each other, they coalesce. 
BS mechanism is thus a Brownian-Coagulation process. 
Tanaka argued for another hydrodynamic growth mechanism involving pressure gradient which is balanced by the thermodynamic force and led to the same exponent $\alpha =1/3$, but a different growth amplitude \cite{tanaka1997new}. 
The growth amplitude ratio for BS and Tanaka is $\simeq 2$. 

The domain growth mechanism is often characterized by the equal-time two-point correlation function $C(r,t)$ which is defined as \cite{puri2009kinetics}
\begin{equation} 
\label{eqcor}
C(r,t)=\langle \psi(\vec r_i,t) \psi(\vec r_j,t) \rangle - 
\langle \psi(\vec r_i,t) \rangle \langle \psi(\vec r_j,t) \rangle,
\end{equation}
Here, $\psi$ is the relevant order parameter and $r=|\vec r_i-\vec r_j|$ represents the distance between two points. 
$C(r,t)$ exhibits data collapse:
\begin{equation}
C(r,t) \equiv \tilde C(r/\ell(t)).
\end{equation}
Here, $\ell(t)$ is the growing lengthscale in the system, viz., the average domain size. 

While the domain growth law provides important information about the kinetics of phase separation, it does not fully characterize the non-equilibrium dynamics. 
In particular, a coarsening system can retain the memory of its age. 
We explore this via the two-time correlation function defined as
\begin{equation}\label{eqacor}
   C_{ag}(t,t_w)=\left\langle\psi(\mathbf r,t_w)\right\rangle
\left\langle\psi(\mathbf r,t)\right\rangle,
\end{equation}
with $\tau=t-t_w$. Here $t_w$ is the waiting time and $t$ is the subsequent observation time. 
If the system were time-translation invariant, the correlation would depend only on the time difference $\tau$. 
However, in an aging system, the correlation depends explicitly on the waiting time $t_w$.
This phenomenon, referred to as \textit{aging}, is a characteristic feature of many slowly evolving non-equilibrium systems \cite{FisherHuse1988, LiuMazenko1991, CorberiLippielloZannetti2006}. 
To the best of our knowledge, aging dynamics has not been studied for bubble coarsening. 

Molecular dynamics (MD) simulations provide a microscopic framework for investigating phase-ordering kinetics \cite{roy2012nucleation, laradji1996molecular} while retaining momentum, mass and hydrodynamics preservation. 
Many-body dissipative particle dynamics simulations demonstrated \cite{Warren2001} that vapor bubbles dispersed in a two-dimensional liquid coarsens through inertial hydrodynamic collapse with $\alpha =2/3$, rather than diffusion-controlled Ostwald ripening. 
More recently, MD simulation study \cite{Parameshwaran2026} of a three-dimensional Lennard-Jones fluid reported a growth exponent $\alpha=1$ consistent with viscous hydrodynamic coarsening. 
These existing studies motivate a systematic reassessment of the microscopic mechanism governing bubble coarsening. 

In the present work, we investigate both domain growth and two-time aging dynamics during vapor bubble coarsening in a three-dimensional Lennard-Jones fluid. 
Apart from the one-time observable, viz., average bubble size, the two-time autocorrelation function $C_{\mathrm{ag}}(t,t_w)$ is explored to characterize the waiting-time dependence of relaxation. 
Together, these measurements connect the growth of spatial structures to the loss of temporal memory. 
Bubble coarsening is particularly interesting because unlike solid domains embedded in a solid matrix, or liquid droplets embedded in a vapor background, here the vapor bubbles grow in a mobile high-density liquid matrix. 
Understanding bubble growth and aging will thus provide important information on the role of background material in the dynamics. 

Rest of the paper is organized as follows: 
In \cref{model}, we explain the model and simulation methodology used. 
\cref{results} present our molecular dynamics simulation results for both single-time and two-time observables. 
Particular focus is on the domain growth law and related growth exponents, two-time autocorrelation function for aging. 
Finally, \cref{conclusions} concludes the paper with a brief summary of the work and presents an outlook.

\section{Model and Methods}\label{model}

We consider a three-dimensional system of $N$ identical particles interacting through the Lennard-Jones potential \cite{roy2016felix}
\begin{equation}
    U_{LJ}=4\varepsilon \Big [\left(\frac{\sigma}{r}\right)^{12}-\left(\frac{\sigma}{r}\right)^{6} \Big ].
\end{equation}
Here, $\varepsilon$ and $\sigma$ define the interaction strength and particle diameter, respectively. 
This interaction is truncated at a cut-off distance $r_c=2.5\sigma$ and shifted to zero at the cutoff. 
The modified potential \cite{sutapa2015} is given by 
\begin{equation} \label {LJmodified}
u(r)=
\begin{cases}
U_{\mathrm{LJ}}(r)-U_{\mathrm{LJ}}(r_c)
-(r-r_c)
\left.\dfrac{dU_{\mathrm{LJ}}}{dr}\right|_{r=r_c},
& r<r_c, \\[6pt]
0, & r\geq r_c.
\end{cases}
\end{equation}
The last term in \cref{LJmodified} ensures the continuity of potential as well as force at $r_c$. 
All particles have same mass $m$ and diameter $\sigma$. 

The simulations are performed in a cubic box of linear dimension $L$, with periodic boundary conditions applied along all three Cartesian directions. 
The simulation box contains $N=\rho L^3$ particles. 
Density is given by $\rho=N\sigma^3/V$, where $V$ is the system volume $V=L^3$. 
We set $m=\sigma=\varepsilon=k_B=1$. 
This fixes the timescale $\tau = m\sigma^2 /\varepsilon$ to unity. 
For this model, the critical temperature $T_c$ \cite{demixing, crctemp} and critical density $\rho_c$ of the vapor-liquid transition were estimated to be $k_B T_c \simeq 0.9\varepsilon$ and $\rho_c \simeq 0.3$, respectively. 

The system is initially equilibrated at a temperature $T_i=5>T_c$, where the fluid is homogeneous. 
Phase separation is initiated by an instantaneous quench to $T_f=0.6T_c$, which lies below the liquid-vapor critical temperature. 
Following the quench, density fluctuations develop and vapor regions form within the liquid phase. 
The system subsequently evolves toward larger characteristic length scales through bubble growth, dissolution, and coalescence. 
Molecular dynamics is performed using a timestep of $\Delta t=[\text{value}]$ using the software package LAMMPS \cite{Plimpton1995}. 
The temperature is controlled using the Nosé-Hoover thermostat \cite{roy2024critical, frenkel2015}. 
For statistical averaging, $20$ independent realizations are considered.

In order to calculate structure characterizing quantities, the fluid is mapped \cite{roy2013dynamics} onto an Ising lattice \cite{func}. 
Particles are moved from their original off-lattice locations to the nearest lattice sites of a simple cubic lattice. 
Next, a filled lattice site is assigned a spin value $S=+1$ and an empty site is assigned a spin value $S=-1$. 
Further, noise is eliminated via a majority spin rule. 
This mapped system is used to calculate domain size and auto-correlations.

\section{Results}\label{results}

\begin{figure}
\centering
\includegraphics*[width=0.5\textwidth]{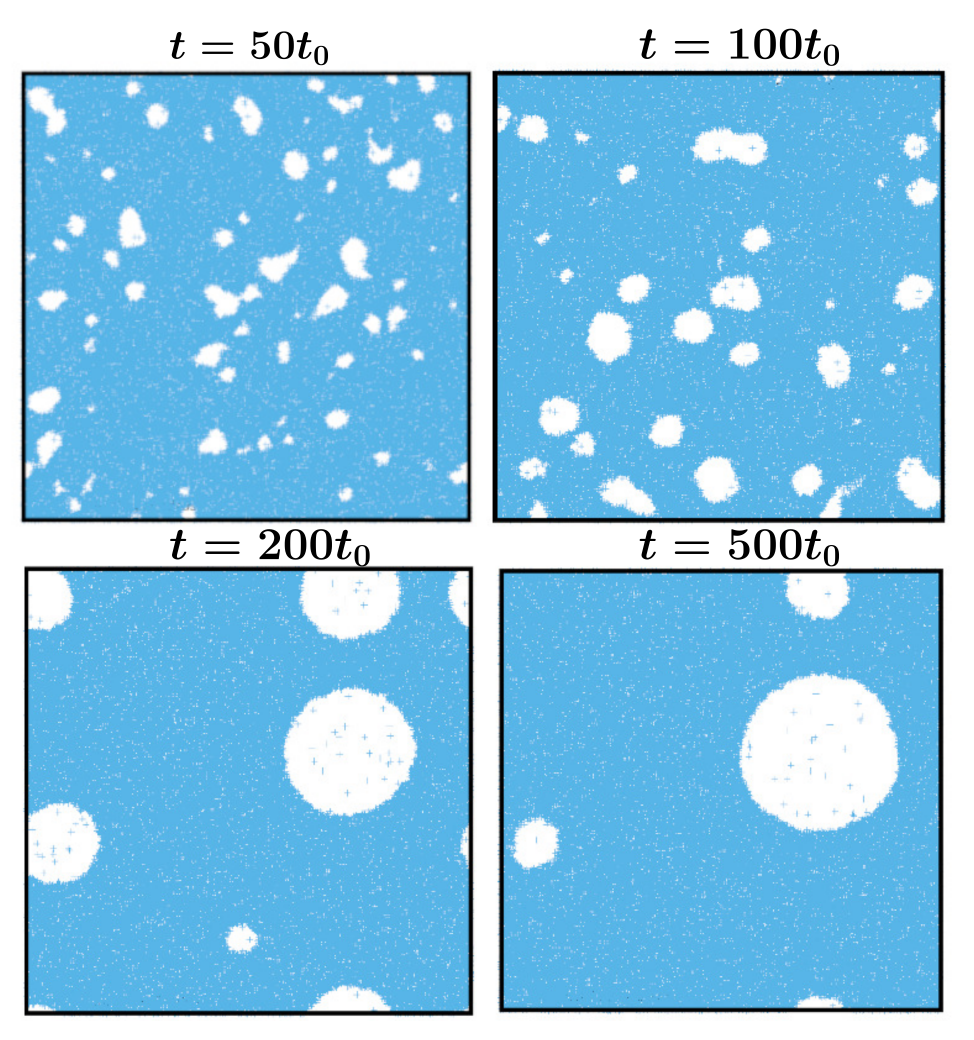}
\caption{Representative two-dimensional cross-sectional snapshots illustrating the temporal evolution of vapor bubbles following a temperature quench from $T_i>T_c$ to $T_f<T_c$. 
The snapshots correspond to the $x$-$y$ plane at fixed $z=100\sigma$, a cubic simulation box of size $L=300 \sigma$, and the quench temperature $k_BT_f=0.6\varepsilon$.
Low-density vapor and high-density liquid phases are shown in white and blue colors, respectively. 
At early times ($t=50t_0$), the morphology is characterized by a large number of relatively small vapor bubbles distributed throughout the liquid phase. 
As time progresses ($t=200t_0$, $500t_0$) the characteristic size of the vapor bubbles increases, accompanied by a systematic decrease in the number of distinct bubbles. 
These are characteristic signatures of `coarsening'.}
\label{fig1}
\end{figure}

\subsection{Bubble Growth Mechanism and Single-Time Observables}

\Cref{fig1} shows representative snapshots illustrating the temporal evolution of the system during phase separation. 
An initially homogeneous system above the critical temperature, $T_c$, is instantaneously quenched below $T_c$. 
For better visualization, two-dimensional cross-sectional slices in the $x$-$y$ plane at a fixed $z=100$ are presented. 
Note that original simulations are performed in three dimensions. 
Since the vapor bubbles grow within the background of the high-density liquid phase, these two-dimensional cross sections provide a clearer understanding of the evolving bubble morphology than the three-dimensional representations. 
In the snapshots, the low-density vapor phase is shown in white, whereas the surrounding high-density liquid phase is represented in blue.
As observed, shortly after the quench, the initially homogeneous system becomes unstable and starts forming bubbles. 
At an early time $t=50t_0$, a large number of relatively small vapor bubbles can be observed throughout the liquid background, where $t_0$ denotes the characteristic time unit of the model (see \Cref{model} for details). 
The morphology at this stage is therefore characterized by a relatively small characteristic length scale and larger number of bubbles.
With increasing time, bubbles merge with each other and grow in size. 
The total number of bubbles keep decreasing with increasing time. 
In particular, at $t=200t_0$ and $500t_0$, the system contains considerably fewer bubbles.
This kind of simultaneous growth of domain size and reduction in the number of domains are signatures of phase separation process. 

\begin{figure}
\centering
\includegraphics*[width=0.5\textwidth]{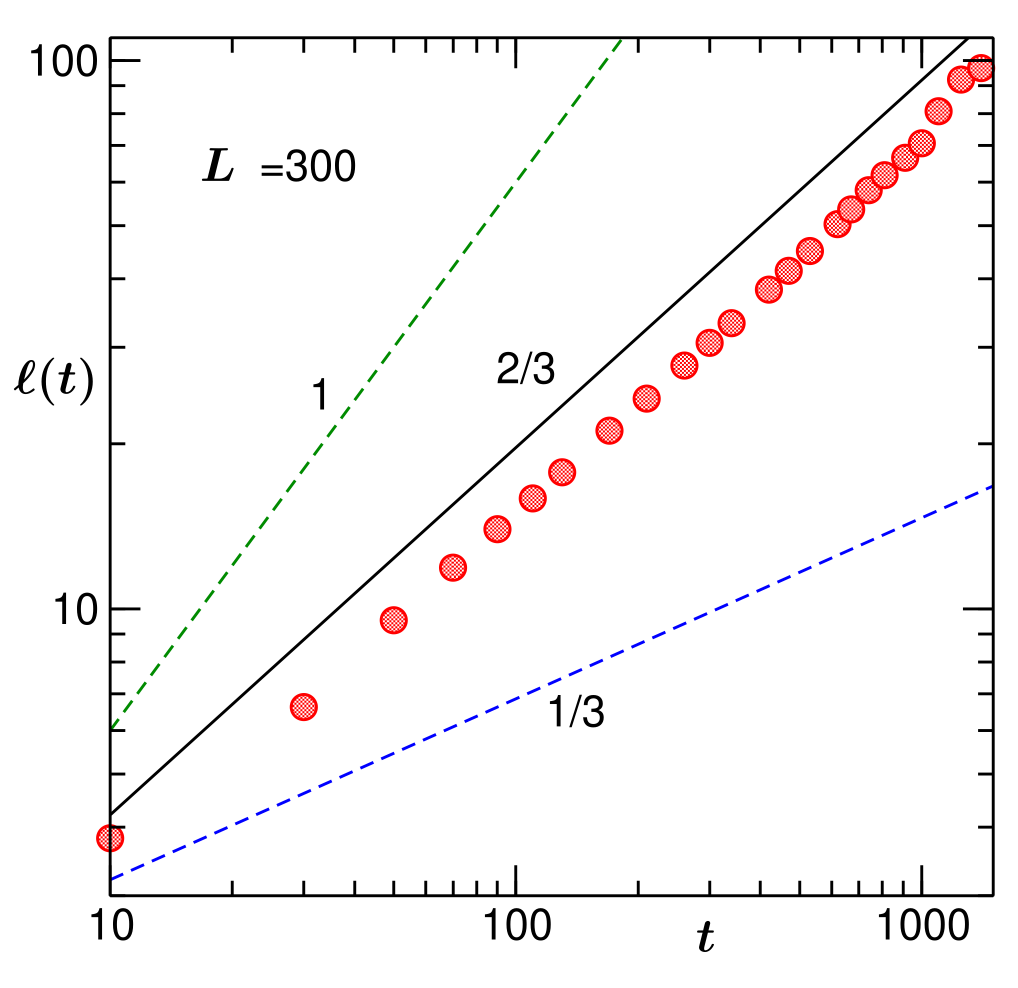}
\caption{Characteristic domain length $\ell(t)$ as a function of time on a log-log scale. The solid black line represents the inertial growth exponent $\alpha=2/3$, while the dashed lines stand for the viscous regime $\alpha=1$ (green) and BS growth exponent $\alpha=1/3$ (blue). Symbols correspond to the MD simulation data from $L=300\sigma$ which corroborates well with $\alpha=2/3$ in the time window $t \simeq (20:700)$.}
\label{fig2}
\end{figure}

The time evolution of the characteristic bubble size, $\ell(t)$, provides a quantitative measure of the coarsening dynamics. 
\Cref{fig2} shows a plot of $\ell(t)$ vs. time $t$, on a double-logarithmic scale. 
Symbols correspond to three-dimensional MD simulation data for system size $L=300\sigma$. 
The solid black line represents the inertial growth exponent $\alpha=2/3$, while the dashed lines stand for the viscous regime $\alpha=1$ (green) and BS growth exponent $\alpha=1/3$ (blue). 
First of all, $\ell(t)$ exhibits a clear algebraic growth over the coarsening regime accessible within the present simulations. 
MD data corroborates well with $\alpha=2/3$ in the time window $t \simeq (20:900)$, rather than with $\alpha=1$ or $1/3$. 
This observation agrees with the hydrodynamic growth exponent reported by Warren \cite{Warren2001} from two-dimensional, particle-based dissipative particle dynamics (DPD) simulations. 

It is important to compare the growth exponent for bubble-growth with that for liquid droplets. 
For droplets, theory predicts a substantially smaller exponent $1/3$ (\cref{BSeq}) \cite{binder1977theory}, which has also been confirmed in MD simulations \cite{sroy2013, roy2013effects}. 
On the other hand, for bubbles we observe $\alpha \simeq 2/3$. 
This difference is striking because for both bubbles and droplets the domain morphology is disconnected in nature unlike the percolating tube-like domains - yet the two systems exhibit distinct growth exponents. 

\begin{figure}
\centering
\includegraphics*[width=0.5\textwidth]{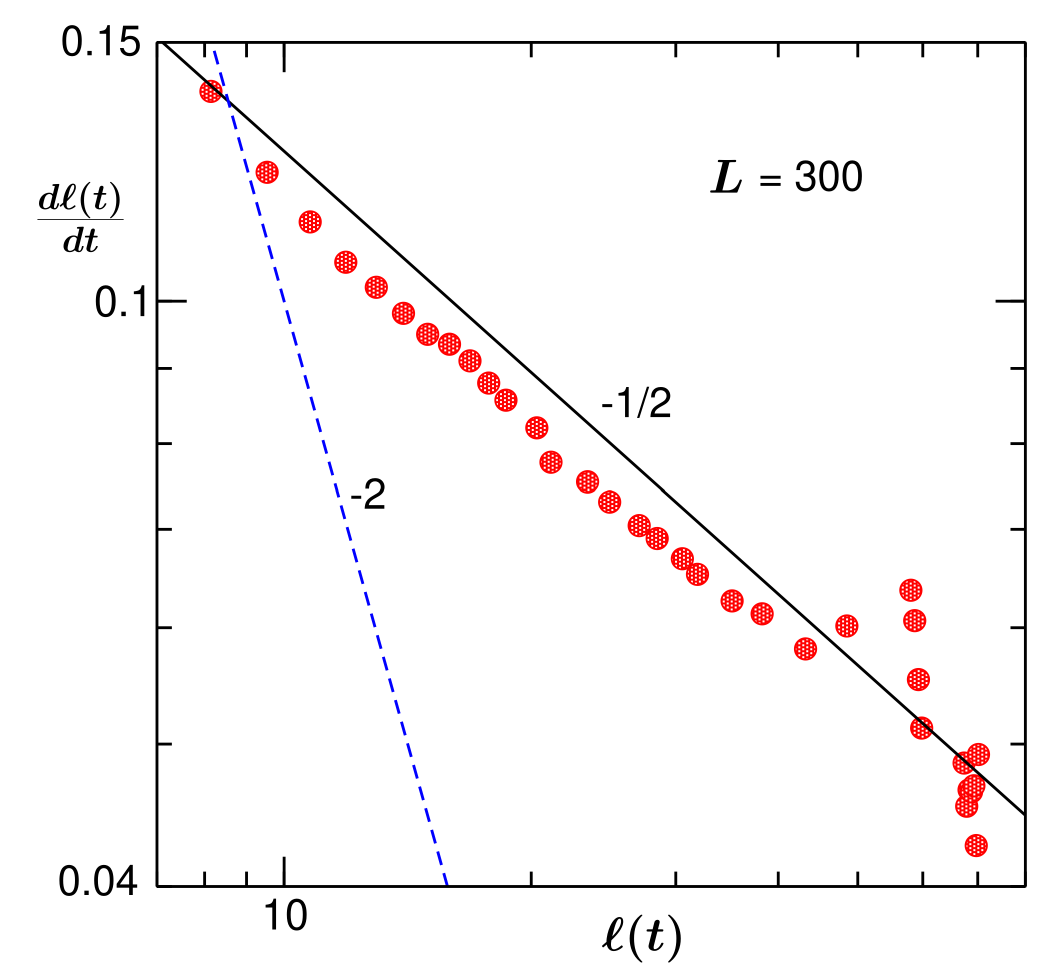}
\caption{Double-logarithmic plot of $d\ell(t)/dt$ vs. $\ell$, where $\ell(t)$ is the average bubble size. 
Symbols refer to the MD simulation data for $L=300\sigma$. 
The solid black line stands for an exponent $\beta =-1/2$ corresponding to the inertial hydrodynamic growth mechanism. 
Definition of $\beta$ is provided in \cref{betaeq}. 
Blue dashed line stands for $\beta=-2$ of Brownian-coagulation mechanism. 
}
\label{fig3}
\end{figure}

To further understand the difference in $\alpha$, we explore the underlying coarsening mechanism. 
Let's consider a domain growth with $\ell(t) \sim t^{\alpha}$. 
This yields 
\begin{equation}\label{betaeq}
    \frac{d \ell(t)}{dt} \sim \ell(t)^{\beta},~\beta=(\alpha-1)/\alpha. 
\end{equation}
LSW mechanism ($\alpha=1/3$) yields $\beta=-2$. 
For inertial hydrodynamic mechanism, $\beta=-1/2$ and for viscous growth mechanism $\beta=0$. 
In \cref{fig3}, we test \cref{betaeq} by plotting $d\ell(t)/dt$ against $\ell(t)$ for the bubble growth dynamics. 
The simulation data (red symbols) for $L=300\sigma$ corroborate well with the inertial mechanism $\beta = -1/2$ (black solid line), rather than the Binder–Stauffer Brownian coagulation mechanism $\beta=-2$ (blue dashed line). 
Further, the MD data clearly do not follow the viscous growth mechanism ($\beta$ = constant). 
These results convincingly support the dominance of hydrodynamic inertial coarsening mechanism during phase separation and growth of vapor bubbles in a liquid background. 

\begin{figure}
\centering
\includegraphics*[width=0.5\textwidth]{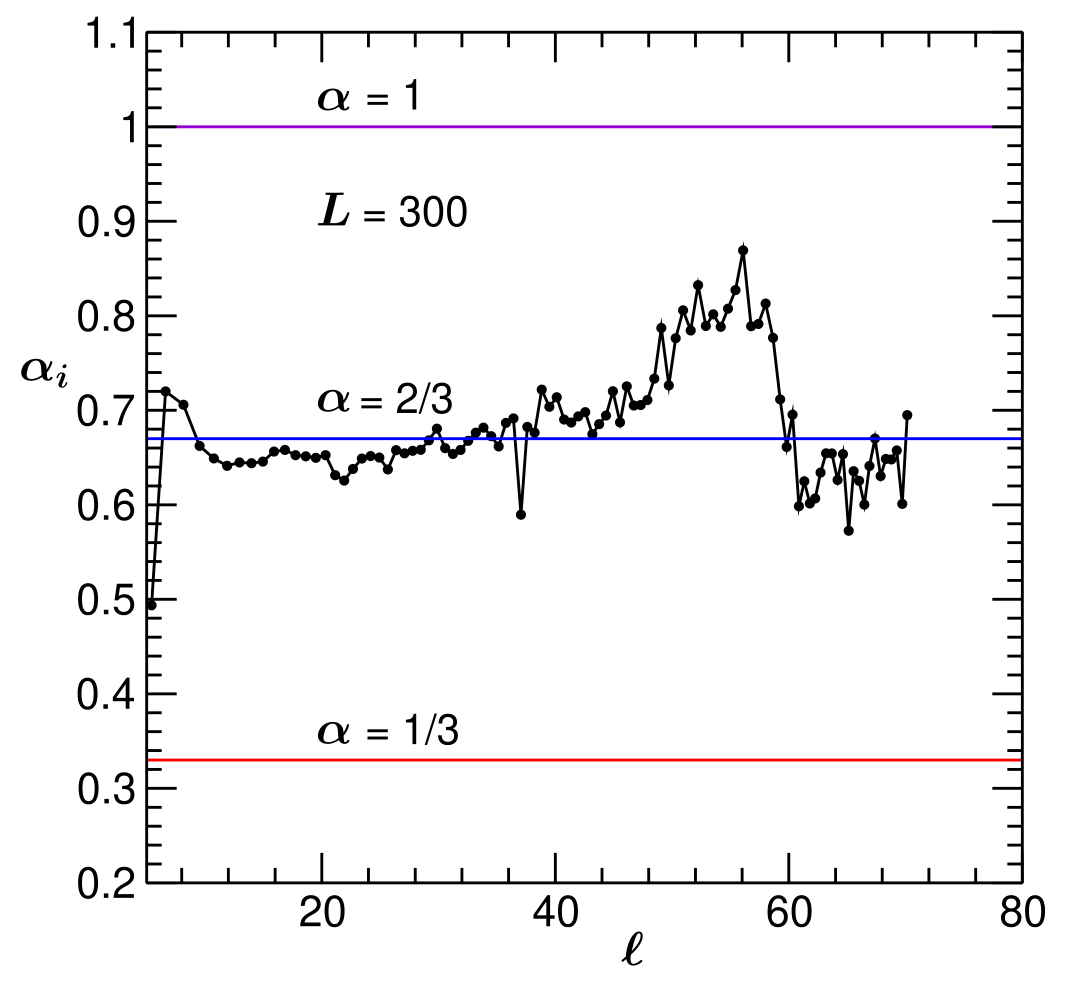}
\caption{Plot of the instantaneous growth exponent $\alpha_i(t)$ vs. $\ell(t)$ for a system size $L=300 \sigma$ (black symbol). 
Domain growth exponents for the viscous ($\alpha=1$), inertial ($2/3$), and the Binder-Stauffer mechaism of Brownian-Coagulation ($1/3$) are also marked by solid lines. 
MD data corroborates well with the inertial growth mechanism $\alpha=2/3$.
}
\label{fig4}
\end{figure}

This observation is further supported by \cref{fig4} in which we plot the time dependence of the instantaneous growth exponent
\begin{equation}
    \alpha_i(t)=\frac{d\log\ell(t)}{d\log t}.
\end{equation}
In \cref{fig4}, for domain sizes $10 \sigma \lesssim \ell \lesssim 30 \sigma$, $\alpha_i$ is consistent with $2/3$ rather than $1/3$ (BS or LS mechanism) or $1$ (viscous growth). 
The larger fluctuations at late times are due to inadequate statistical averaging.

An important aspect of the present study is the use of a substantially larger simulation system, with linear dimension $L=300 \sigma$, which allows the power-law regime to be examined over an extended temporal window while reducing the influence of finite-size effects on the measured growth behavior.

Domain morphology is central to the interpretation of the growth exponent $\alpha$. 
Siggia’s viscous mechanism \cite{edsiggia1979} assumes connected tube-like domains through which fluid flows due to surface-tension-induced pressure differences. 
For our system the growing vapor bubbles are spatially disconnected and hence the tube-like transport pathway does not apply.
Hydrodynamic flow can still occur through the continuous background liquid surrounding the bubbles. 
Brownian translational diffusion followed by binary collision/coalescence (BS) acts as the driving mechanism for the growth of liquid droplets during off-critical phase separation \cite{binder1977theory}. 
BS mechanism incorporates hydrodynamics. 
Results presented in \cref{fig2}, \cref{fig3} already demonstrated that BS is not the dominant mechanism for growth of bubbles. 
Instead, inertia-dominated hydrodynamic coarsening is the main driving mechanism. 
This may be explained in terms of the density of the background phase. 
Liquid droplets grow within a low-density vapor matrix and exhibit a smaller growth exponent $1/3$, whereas vapor bubbles grow within a dense liquid matrix and exhibit a higher exponent $2/3$. 
The surrounding fluid thus clearly plays a crucial role in the domain growth mechanism.

\subsection{Aging Dynamics and Two-Time Quantities}

\begin{figure}
\centering
\includegraphics*[width=0.5\textwidth]{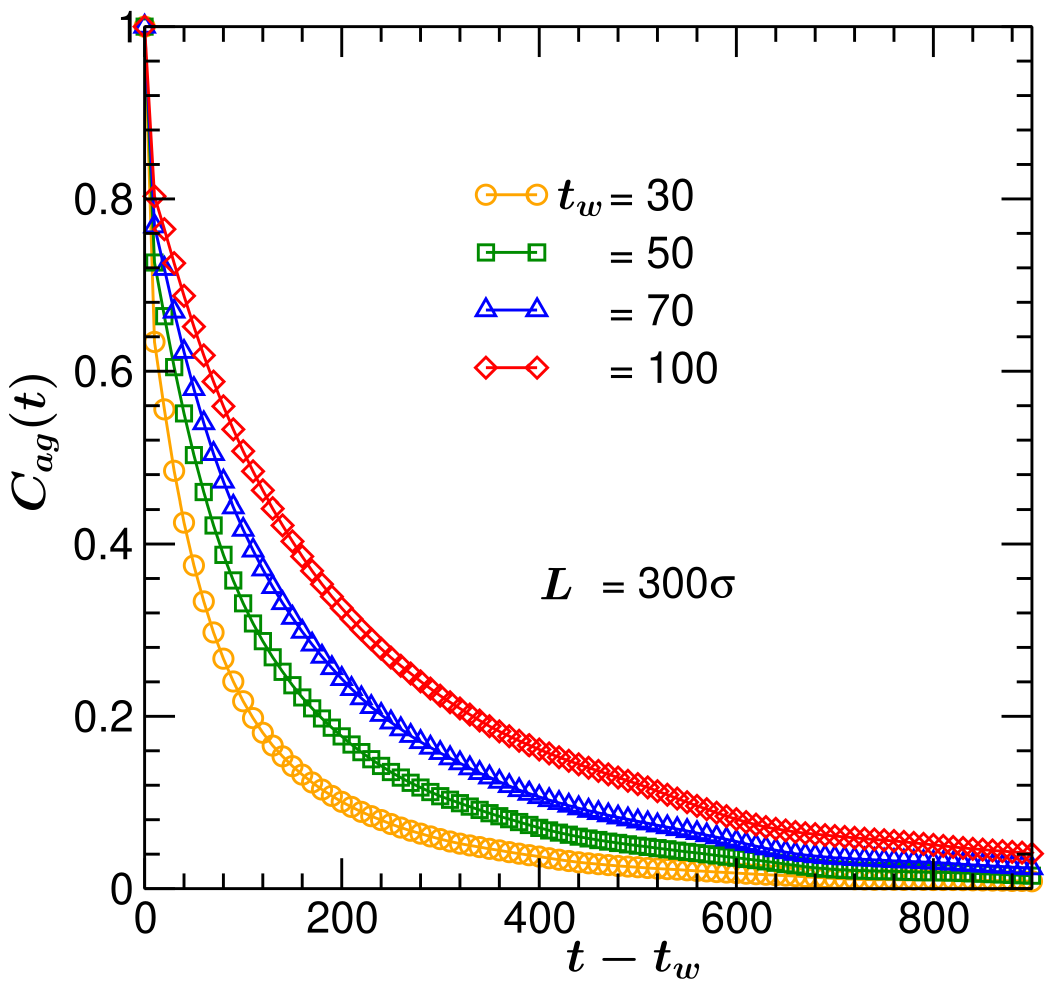}
\caption{Two-time autocorrelation function $C(t,t_w)$ vs. $(t-t_w)$, for four distinct values of $t_w$. 
The slower decay at larger $t_w$ demonstrates aging.
}
\label{fig5}
\end{figure}

The average domain size $\ell(t)$ characterizes the morphology and growth dynamics at a single time. 
Two-time quantities provide deeper insights into the non-equilibrium dynamics of the system. 
We calculate the two-time autocorrelation function $C_{ag}(t,t_w)$ (\cref{eqacor}) which is a measure of how a system retains a memory of its state at an earlier time. 
In \cref{fig5}, $C_{ag}(t,t_w)$ is plotted as a function of time window $(t-t_w)$, for four distinct values of the waiting time $t_w$. 
Clearly, a systematic dependence on $t_w$ is observed. 
$C_{ag}(t,t_w)$ decays more slowly when the waiting time $t_w$ is increased. 
Thus, an older system evolves more slowly than a younger system. 
If time-translation invariance was held, the correlation would have depended only on the elapsed time $(t-t_w)$ and not on $t_w$. 
However, in \cref{fig5}, $C_{ag}(t,t_w)$ depends on $t_w$: this is a signature of `aging' \cite{sutapa2019-aging}. 

The physical origin of this behavior can be understood from the continuously increasing characteristic length $\ell(t)$. 
At early times, the system contains many small bubbles and the morphology changes relatively rapidly. 
As coarsening proceeds, the number of bubbles decreases and the characteristic length increases. 
The subsequent evolution therefore involves progressively larger structures and occurs on increasingly long time scales. 
Relaxation gets slower. 

\begin{figure}
\centering
\includegraphics*[width=0.5\textwidth]{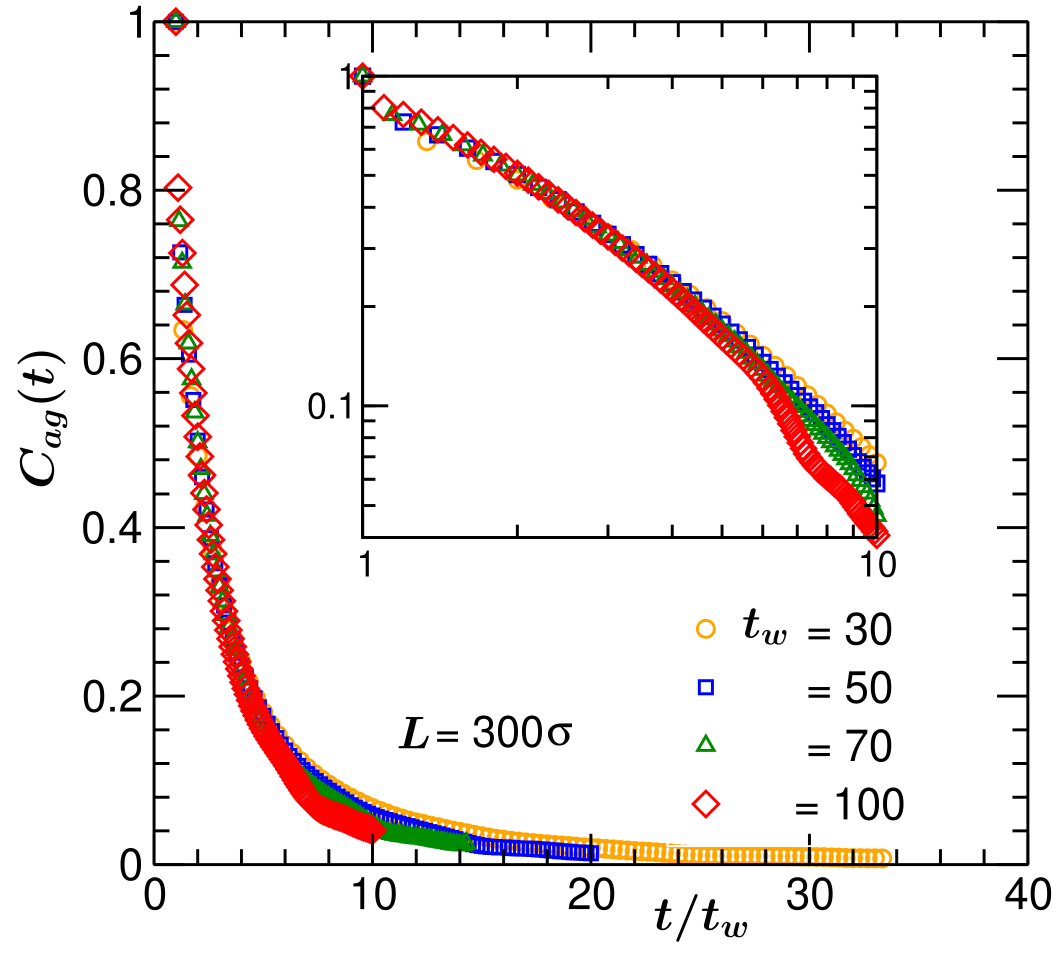}
\caption{Two-time autocorrelation function $C_{ag}(t,t_w)$ vs. scaled $t/t_w$, for four values of $t_w$. Inset: same as main frame, but on a double-logarithmic scale. 
}
\label{fig6}
\end{figure}

Next, in \cref{fig6}, we plot $C_{ag}(t,t_w)$ vs. $t/t_w$, for the same values of $t_w$. 
A nice data collapse is observed which indicates that the relaxation time grows as $\sim t_w$. 
The larger fluctuations for larger $x$ indicate the requirement of better statistical averaging. 
The inset shows same data, but on a double-logarithmic scale. 
Slight bending of the curves indicate that the $C_{ag}(t,t_w)$ scaling function is not a pure power-law; some exponential corrections as observed in \cite{Das-Paul2026-aging, Midya2020aging} might be present.

\begin{figure}
\centering
\includegraphics*[width=0.5\textwidth]{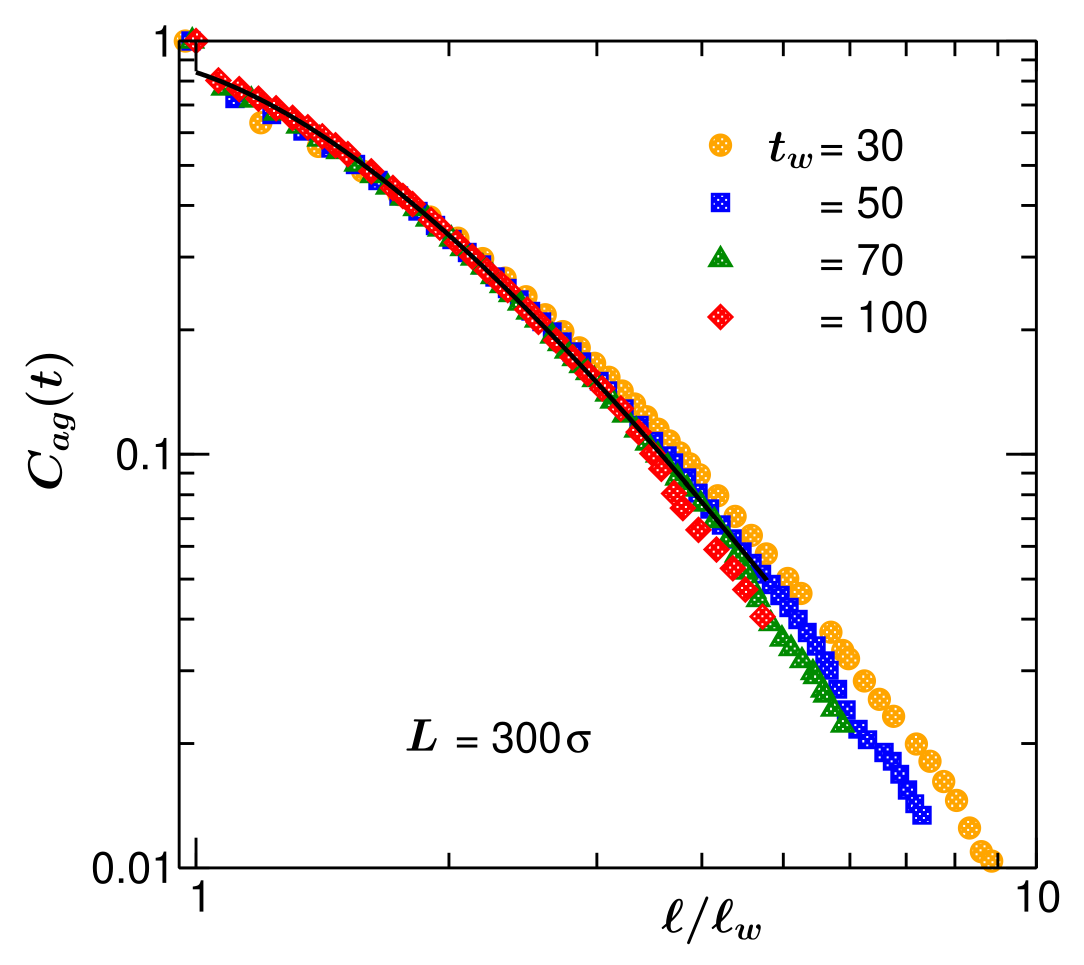}
\caption{Scaling plot of $C_{ag}(t,t_w)$ as a function of $\ell(t)/\ell(t_w)$. The curves for different waiting times $t_w$ collapse nicely onto a master curve marked by black solid line. 
This collapse indicates that the lengthscale $\ell(t)$ controls the memory of the phase separating fluid. 
}
\label{fig7}
\end{figure}

The two-time correlation can be further analyzed using the characteristic domain length $\ell(t)$. 
In \cref{fig7}, we plot $C_{ag}(t,t_w)$ vs. $\ell(t)/ \ell(t_w)$.  
Note that the x-axis is dimensionless. 
Data for different waiting times $t_w$ almost collapse onto a common scaling function $\mathcal F$:

\begin{equation}
    C_{ag}(t,t_w) = \mathcal F \Big ( \frac{\ell(t)}{\ell(t_w)}\Big).
\end{equation}
This collapse demonstrates the aging dynamics is governed by the evolving characteristic length scale $\ell(t)$.
The absolute time elapsed after the quench is not the most appropriate variable for describing the two-time dynamics. 
Instead, the relevant measure of dynamical evolution is the amount by which the domain length has increased relative to its value at the waiting time. 

The exact functional form of $\mathcal F$ is rather interesting. 
Bending of the curves in \cref{fig7} indicate that $\mathcal F$ is not purely algebraic. 
In fact, they agree well with the function \cite{Midya2014Aging}
\begin{equation}
    \mathcal F \Big(x=\frac{\ell(t)}{\ell(t_w)} \Big)= A x^{-\lambda} \text{exp} \Big(-C/x \Big),
\end{equation}
where, $A$, $C$ are positive constants and $\lambda \simeq 3$. 
For droplet growth during phase separation, a value of $\lambda \simeq 6.6$ was observed \cite{sutapa2019-aging}. 
Clearly, even though both droplets and bubbles are morphology-wise disconnected in nature, they do not yield the same value of $\lambda$. 
This might be attributed to the difference in background matrix density. 

The scaling collapse provides a direct connection between the one-time coarsening dynamics and the two-time aging dynamics. 
The same length scale $\ell(t)$ that controls the spatial self-similarity of the phase-separated morphology also controls the temporal memory of the system.

\section{Conclusions}\label{conclusions}

We have investigated both the domain growth mechanism and aging dynamics during phase separation of vapor bubbles in a three-dimensional Lennard-Jones model fluid using large-scale molecular dynamics (MD) simulations. 
Following an instantaneous quench from a high temperature ($>T_c$) to a temperature inside the co-existence region, vapor bubbles form in the liquid background and they grow with time. 
During the non-equilibrium evolution, while the bubbles grow their total number decreases. 
Nos\'{e}--Hoover thermostat \cite{martyna1992, frenkel2015} is applied to maintain the fluid at the desired temperature. 
This thermostat is known to preserve hydrodynamics well. 

Unlike solid droplets embedded in a solid matrix for a binary mixture, or liquid droplets embedded in a vapor background, 
here the vapor bubbles are floating in a mobile liquid matrix. 
The density of the background phase, presence of hydrodynamic flow are distinct in these three scenarios. 
It is thus noteworthy to explore the differences of the domain growth mechanism for bubbles with other systems. 

Our MD simulation results demonstrate that the average bubble size $\ell(t)$ grows as a function of time with an exponent $\alpha \simeq 2/3$ which is characteristic of inertial hydrodynamic coarsening. 
Our simulation data also negate the possibility of the Binder-Staufer (BS) growth exponent $1/3$ as well as the diffusion-controlled coarsening leading to $\alpha=1/3$. 
Viscous hydrodynamic growth does not appear to be the driving mechanism. 

For the growth of liquid droplets, theory predicts a substantially smaller exponent $\alpha=1/3$ (\cref{BSeq}) \cite{binder1977theory}, which was also confirmed in MD simulations \cite{sroy2013, roy2013effects}. 
On the other hand, for bubbles a distinctly different exponent is observed. 
This difference is striking because for both bubbles and droplets the domains are of disconnected morphology -- yet the two systems exhibit distinct growth exponents. 

We attribute the observed growth exponent to the domain morphology. Siggia’s viscous mechanism \cite{edsiggia1979} assumes connected tube-like domains and hence this mechanism is not observed in our system. 
For liquid droplets growing in the low-density vapor matrix Brownian motion, collision and coalescence becomes dominant and leads to $\alpha=1/3$. 
For bubble growth, the background liquid matrix has a density higher than the bubbles and allows for hydrodynamic flow through it and leads to a much faster growth of bubbles. 
An important conclusion from this study is thus the surrounding fluid thus clearly plays a crucial role in the domain growth mechanism.

Beyond the conventional analysis of the single-time observables,  we have also investigated the two-time autocorrelation function $C_{ag}(t,t_w)$, where $t_w$ and $t$ are the waiting and observation times, respectively. 
Understanding $C_{ag}(t,t_w)$ is particularly important as it sheds light on the memory of the system. 
Evolution between two times can depend not only the observation time, but also on the waiting time. 
As observed in our study, $C_{ag}(t,t_w)$ exhibits slower decay at later times. 
This is a characteristic property of `aging'. 
The aging curves collapse when expressed in terms of the ratio of the characteristic lengths $\ell(t)/\ell(t_w)$. 
The evolving domain length thus provides an appropriate scaling variable for the aging dynamics. 
The master curve corroborates well with an exponential correction to a power-law behavior: $Ax^{-\lambda} exp(-C/x)$, where $A$ and $C$ are constants. 
For bubble growth, $\lambda$ is estimated to be $\simeq 3$. 
Interestingly, for droplet growth $\lambda \simeq 6.6$ \cite{sutapa2019-aging}. 
Even though both droplets and bubbles are disconnected morphology they yield distinctly different $\lambda$.

A particularly important implication is that the growth of the characteristic lengthscale, memory effects, and relaxation of the system strongly depend on the background matrix for fluid phase separation. 
We hope these findings will motivate further studies to address universality of aging exponents across various fluids of different densities, different temperature quenches. 
The latter will provide a better perspective on the coupled growth and memory effects associated with disconnecetd droplet morphology.

\vskip 1.0cm
Acknowledgement: SR thanks financial support from BITS Pilani, India, through NFSG grant no: N5/25/1053.


\bibliography{phasesep}
\end{document}